\documentstyle[prd,aps,eqsecnum]{revtex}

\input epsf.tex
\tighten
\begin{document}
\draft
\twocolumn[\hsize\textwidth\columnwidth\hsize\csname 
@twocolumnfalse\endcsname

\title{Cold dark matter identification: Diurnal modulation reexamined}

\author{F. Hasenbalg$^1$, D. Abriola$^2$, J.I. Collar$^3$, D.E.  Di
Gregorio$^2$, A.O. Gattone$^2$, H. Huck$^2$, D. Tomasi$^2$, and I.
Urteaga$^2$}

\address{$^1$ Department of Physics and Astronomy, University of South
Carolina, Columbia, SC 29208 \\ $^2$ Departamento de F\'{\i}sica,
Comisi\'on Nacional de Energ\'{\i}a At\'omica, Av. del Libertador 8250,
1429 Buenos Aires, Argentina\\ $^3$ PPE Division, CERN, Geneve 23,
CH-1211, Switzerland}

\date{Received 4 December 1996}
\maketitle

\begin{abstract}
We report on new estimates of the modulation expected in semiconductor
detectors due to eclipsing of dark matter particles in the Earth.  We
reevaluate the theoretical modulation significances and discuss the
differences found with previous calculations. We find that a
significantly larger statistics than previously estimated is needed to
achieve the same level of sensitivity in the modulated signal.
\end{abstract}

\vskip2pc]

\section{Introduction}

For the past ten years or so semiconductor detectors (germanium and
silicon) have been used to place bounds on masses and cross sections of
cold dark matter candidates~\cite{Ahlen,caldwell,mosca,avignone}.  In
these devices massive particles in the halo of our Galaxy, with
interactions weaker than electromagnetic, can be detected by measuring
the recoil energy produced by their elastic scattering off the detector
nuclei.

To discriminate the signature of these weakly interacting massive
particles (WIMP's) from the background, the first idea advanced
\cite{drukier} was to look for annual modulations of the detection
rates originating in the orbital motion of the Earth around the Sun.
Due to this, the average WIMP velocity relative to the Earth does not
remain constant in time, giving rise to a variation, between June and
December, of the order of 4\% to 6\%~\cite{drukier}.

In 1992, Collar and Avignone~\cite{collarPLB} put forward the idea of
looking for daily---as opposed to annual---modulations in counting
rates and energy spectra. The argument was based on the fact that the
Earth, during its daily rotation, could act as a shield in front of the
detector eclipsing the dark matter wind traversing the planet. For
candidates with scalar or vector interactions (axial-vector
interactions with the Earth constituents are small) this would modify
fluxes and velocity distributions, thus affecting counting rates. The
theoretical estimates of this effect were obtained by Monte Carlo
calculations (MC) by these authors~\cite{juanD} and indicated that for
a suitably located detector the modulation could be of the order of 2
to 10\% for the range of masses for which semiconductor detectors are
more sensitive (10~GeV to 10~TeV) and for couplings of the order of the
weak coupling constant~\cite{juanD}. Either type of search, annual or
diurnal, requires the signal to be collected in the form of ``event by
event;" namely, each event must be time tagged.

Because of its shorter period, daily modulation has the advantage of
placing less stringent requirements on the stability of the detector
and its associated electronics. The first results ever published of a
search for daily modulation appeared some time
ago~\cite{garciaD,Sarsa,garcia} and were performed with a Ge detector
in the Canfranc tunnel. The experiment, run by the USC/PNNL/UZ
(University of South Carolina, Pacific Northwest National Laboratories,
Zaragoza) collaboration was located, however, in a geographical
location which did not favor this sort of search. To improve this, a
new site was sought and eventually found in Sierra Grande, Argentina,
where a laboratory was set up and a 1.033~kg Ge detector installed in
mid 1994. The TANDAR group, with support of the rest of the
collaboration, has been running the experiment since then and the first
analyses of the data were published elsewhere~\cite{taup95}.

To get acquainted with the details involved in the calculation and to
carry out an independent check of the previous theoretical estimates, a
new MC code was developed \cite{hasenbalg}. The new results indicate
that, for masses larger than $\sim$ 1~TeV (and couplings as large as 10
times the weak coupling constant) the modulation of the signal is
predicted to be less than 1\%. This is at variance with the original
calculations and implies that a larger amount of data than previously
estimated (more than 20~kg~yr for a modulation of $\approx$ 0.5\% as
opposed to 2~kg~yr) is needed for the sensitivity of the diurnal
modulation method to exceed that of the conventional
signal-to-noise~\cite{Ahlen,garciaD}.

The aim of this paper is to report on new estimates, to evaluate the
expected modulation significances, and to discuss the source of the
discrepancy with respect to the original calculation. The next section
describes the input to the calculation and gives a general idea of its
contents. In the last section we give an account of the results and we
discuss the differences with previous reports. Finally, we present some
conclusions.

\section{Expected Daily Modulation}

For an incoming WIMP of mass $m_{\chi}$ and velocity $v$ ($\beta =
v/c$) the total counting rate in the recoil-energy interval $T$,
$T+dT$ of the Ge nuclei, is given by

\begin{equation}
\frac{dN}{dTdt}(T) = N_A \left(\frac{\rho_{halo}}{m_\chi} \right)
     \int_{v_{min}}^{v_{max}} g'(v) v \frac{d\sigma}{dT}(v,T) dv,
     \label{rate}
\end{equation}

\vspace{-1.4cm}

\begin{figure}
\centering
\epsfxsize=7.5truecm
\epsffile{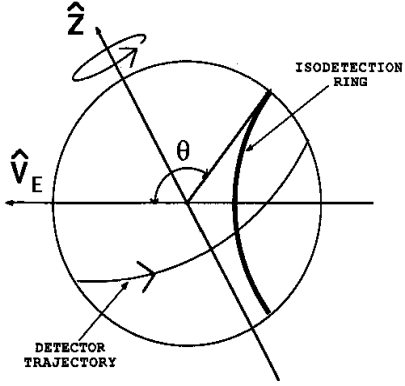}
\caption{Relevant directions in the scattering of WIMP's by the Earth
as it travels around the galactic center with velocity $\vec{V}_{E}$.
The angle $\theta$ defines {\em isodetection rings} and a zenith
angle.\label{fig1}}
\end{figure}

\noindent
where $N_A$ is the total number of atoms in the detector,
$\rho_{halo}/m_\chi$ is the number density of WIMP's ($\rho_{halo} =
0.3$~GeV~cm$^{-3}$) and $g'(v)$ is the Maxwell-Boltzmann velocity
distribution of the WIMP's for an observer on the Earth:
namely~\cite{Freese},

\begin{equation}
  g'(v)\, dv \;=\; \frac{4}{\sqrt{ \pi}}\, \frac{\chi^3}{v}\,
      e^{-(\chi^2 + \eta^2)}\, \frac{\sinh(2 \chi \eta)}{2 \chi \eta} \, dv,
      \label{MB}
\end{equation}
where
\begin{equation}
 \chi^2 \;=\; \frac{3 v^2}{2 \sigma_v^2}, \hspace{0.7cm} 
     \eta^2 \; =\; \frac{3 V_{E}^2}{2 \sigma_v^2} \;.
\end{equation}
In the previous equation, $V_{E}$ is the velocity of the Earth with
respect to the galactic rest frame ($V_{E}$~$\sim$~230~km~s$^{-1}$),
$\sigma_{v}$ is the velocity dispersion of the WIMP's in the galactic
halo ($\sigma_{v}$ is typically 270~km~s$^{-1}$~\cite{Freese}), and the
distribution (\ref{MB}) is normalized to one. In Eq. (\ref{rate}),
$v_{max}$ is the maximum velocity  

\vspace{-0.7cm}

\begin{figure}[h]
\centering
\epsfxsize=9.0truecm
\epsfysize=6.8truecm
\epsffile{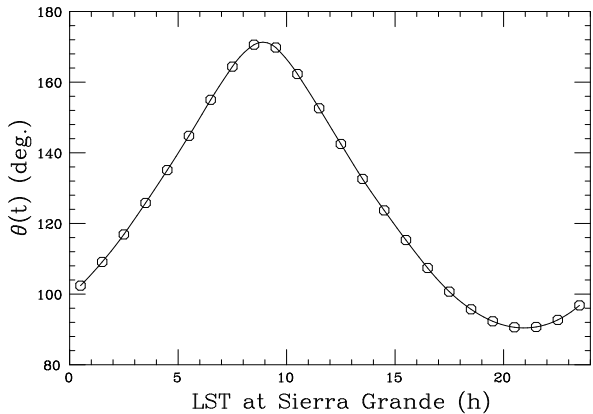}
\caption{Average variation of $\theta(t)$ vs $LST$ at Sierra Grande
throughout the year of 1994. The coordinates of Sierra Grande are
41$^{\rm o}$~40$'$~S, 65$^{\rm o}$~23$'$~W.\label{fig2}}
\end{figure}

\vspace{-0.7cm}

\begin{figure}
\centering
\epsfxsize=9.0truecm
\epsfysize=6.8truecm
\epsffile{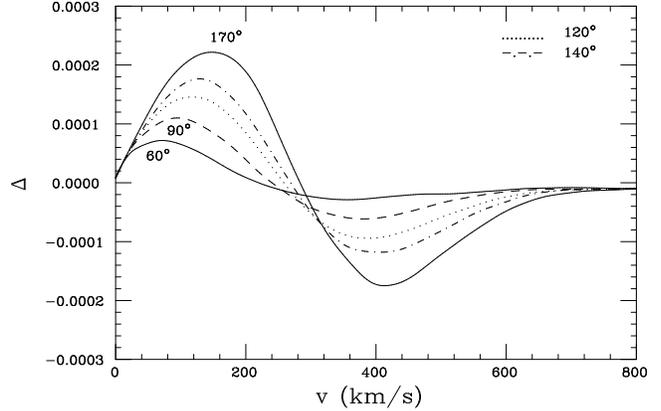}
\caption{Difference between velocity distributions of WIMP's of mass
$M_X$ = 50~GeV for several angles $\theta$ and the original
Maxwell-Boltzmann distribution, $\Delta = P_{MC}(v,\theta)dv - P_{MB}(v)
dv$. Here an interaction strength $g=10$ was assumed.\label{fig3}}
\end{figure}

\noindent
of the WIMP's ($v_{max} = V_{E} + v_{esc}$, where $v_{esc}$ is the
escape velocity from our galaxy, 570~km~s$^{-1}$) and $v_{min}$ is the
minimum velocity of a WIMP necessary to contribute to a particular
energy of the recoil spectrum, $T_{min}$. If $\mu$ is the reduced mass
of the WIMP-nucleus system and $m_N$ is the mass of the recoiling
nucleus, then

\begin{equation}
 v_{min}\; =\; \sqrt{m_N T_{min}/2 \mu^2} \, c \;.
\end{equation}

In Eq. (\ref{rate}) the differential cross section for the scattering
of a WIMP off a spin zero ($J=0$) nucleus in the detector assuming the
exchange of $Z_0$-like vector boson ($G_W^2 =g\, G_F^2$; $G_F =
1.16639 \times 10^{-5}$~GeV$^{-2}$, the Fermi coupling constant) is of
the form~\cite{Ira}

\vspace{0.2cm}

\begin{figure}
\centering
\epsfxsize=9.0truecm
\epsfysize=6.8truecm
\epsffile{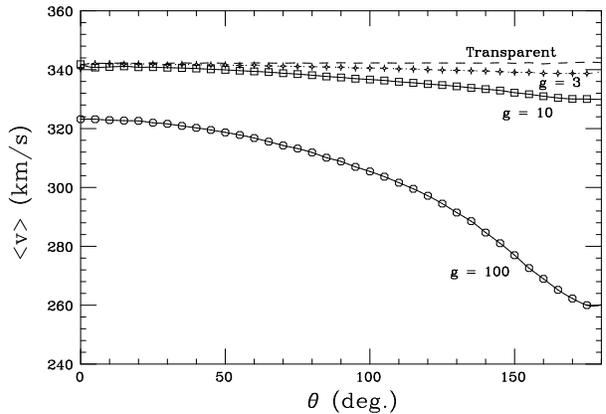}
\caption{Average velocity of WIMP's of mass $M_X$=50~GeV vs.
$\theta$. Three cases $g=3$, $g=10$, and $g=100$ are compared to
the average velocity of WIMP's going through a ``transparent''
Earth.\label{fig4}}
\end{figure}

\vspace{-0.7cm}

\begin{figure}
\centering
\epsfxsize=9.0truecm
\epsfysize=6.8truecm
\epsffile{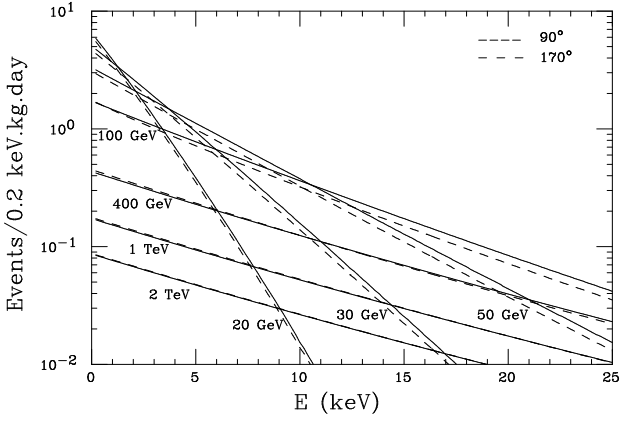}
\caption{Energy spectra for several masses at two different angles,
$\theta = 90^{\rm o}$ (day) and $\theta = 170^{\rm o}$ (night). An
interaction strength $g = 10$ was assumed.\label{fig5}}
\end{figure}

\begin{equation}
\frac{d\sigma}{dT}(\beta , T) = \frac{G_{W}^2}{8\pi} 
  \left[ N - Z ( 1 - 4 \sin^2 \theta_w) \right]^2 
  \frac{m_N}{2 \beta^2}\, e^{-\frac{2m_N T R^2}{3(\hbar c)^2}},
\label{cross}
\end{equation}

\noindent
where $R = 1.2 A^{1/3}$~(fm) is the nuclear radius, $\theta_{w}$ is
the weak mixing angle, and the exponential factor in Eq. (\ref{cross})
takes into account the loss of coherence of the interaction.  The
calculated recoil spectrum must also be convoluted with a function
accounting for the efficiency of the recoiling process generating an
ionization signal.  In our case this function was taken from
Ref.~\cite{Ahlen}.

Estimations of a diurnal modulation involve considering the relation
between the rotational axis of the Earth and its velocity (see
Fig.~\ref{fig1}). The velocity of the Earth, $\vec{V}_{E}$, through the
Galaxy defines an axis of symmetry around which, the dispersion of
WIMP's has azimuthal symmetry. The position vector of a given detector
(with the origin in the center of the planet) and $\vec{V}_{E}$ define
an angle, $\theta$, that plays the role of a zenith angle.  Thus, at
$\theta=0^{\rm o}$ the detector is maximally exposed to the ``WIMP's
wind," whereas at $\theta=180^{\rm o}$ this exposure is diminished by
the bulk of the Earth's mass. The angle $\theta$ also defines {\em
isodetection rings}, or rings of constant flux due to the symmetry of
the problem. To each angle $\theta$ we can associate a time in some
system.  Particularly, the relation between the local sidereal time
(LST) and the isodetection angles $\theta$ crossed by the Sierra Grande
detector during 1994 is shown in Fig.~\ref{fig2}. A detectable
modulation effect can be quantitatively predicted provided there is a
significant change with $\theta$ in the total counting rate and/or
spectral shape.

The purpose of the MC code is to track all the particles and calculate
their number density per unit area and their velocity distributions at
the surface of the Earth, for all isodetection rings. The parameters
and model of Earth are similar to those used in Ref.~\cite{juanD}. A
typical outcome of the Monte Carlo can be seen in Fig.~\ref{fig3} which

\vspace{-0.7cm}

\begin{figure}
\centering
\epsfxsize=9.0truecm
\epsfysize=6.8truecm
\epsffile{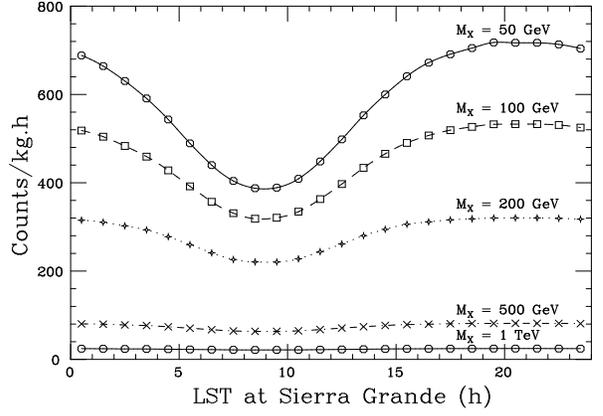}
\caption{Expected rate of events between $E_{min}$ and 50~keV for
WIMP's of different masses as a function of local sidereal time (LST)
at Sierra Grande.  An interaction strength of $g = 100$ was assumed to
make the diurnal modulation effect noticeable.\label{fig6}}
\end{figure}

\noindent
shows how the velocity distributions, of WIMP's of 50~GeV mass and $g =
10$ vary for different angles. The parameter $\Delta$ depicted in the
figure, is defined as the difference between the velocity distribution
of the WIMP's for a given angle $\theta$ and the original
Maxwell-Boltzmann distribution as given by Eq. (\ref{MB}), [$\Delta =
P_{MC}(v,\theta)dv \, -\,P_{MB}(v)dv$].  The distributions are all
normalized to unity, since the number of particles crossing the area
element $dA = 2\pi \sin\theta d\theta$ are not the same.  Note that
the elastic scattering with nuclei of the Earth has the effect of
depleting the region of particles with high velocities and increasing
the number of particles with lower speeds.

A parameter that summarizes the properties of the velocity
distributions is the average velocity distribution of incoming (i.e.,
penetrating the Earth) plus outgoing (exiting the Earth) WIMP's,
$<v>(\theta)$. Figure~\ref{fig4} shows

\vspace{-0.5cm}

\begin{figure}
\centering
\epsfxsize=9.0truecm
\epsfysize=6.8truecm
\epsffile{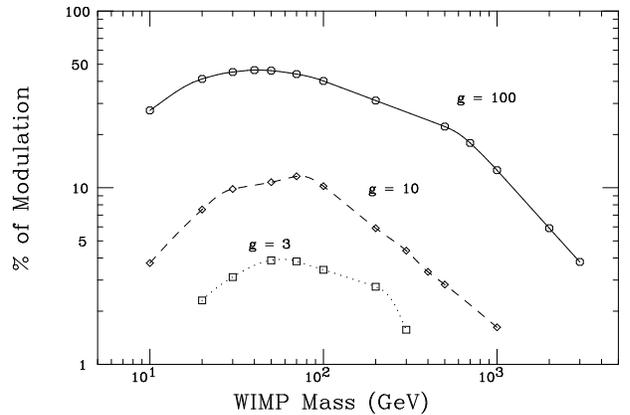}
\caption{Percentage of modulation, $\delta$, as a function of WIMP
mass for different interaction strengths.\label{fig7}}
\end{figure}

\newpage

\widetext

\begin{table}

\caption{Predicted diurnal modulation, $\delta$, for several coupling
constants. $R_{max}$ and $R_{min}$ are the total counts expected in the
energy interval from $E_{min}$ to 50~keV at maximum and minimum
respectively, divided by the energy interval.}
\label{t1}

\vspace{.3cm}


\begin{tabular}{c c c c c c c} 
WIMP & LST     & LST     & $E_{min}$  & $R_{max}$ & $R_{min}$ & $\delta$\\
mass & at max. & at min. & (keV)& (counts/keV~kg~d)&(counts/keV~kg~d)&(\%) \\
(GeV)&   (h)   &    (h)  &            &     &    &           \\
\hline 
\multicolumn{7}{c}{$g = 3$} \\ \hline
20   &  17.5  &  9.5  &  0 &  7.69   &  7.51  &  2.30   \\
30   &  20.5  &  9.5  &  0 & 10.16   &  9.84  &  3.12   \\
50   &  22.5  &  9.5  &  0 & 10.55   & 10.14  &  3.88   \\
70   &  20.5  &  9.5  & 2.5&  6.95   &  6.68  &  3.83   \\
100  &  20.5  &  8.5  &  3 &  5.62   &  5.43  &  3.43   \\
200  &  20.5  &  8.5  & 10 &  1.73   &  1.68  &  2.75   \\
300  &  19.5  &  8.5  & 15 &  0.83   &  0.82  &  1.57   \\
\hline
\multicolumn{7}{c}{$g = 10$}  \\ \hline
10   &  17.5  &  8.5  &  0 &   6.04   &  5.82   &  3.74   \\
20   &  22.5  &  9.5  &  0 &  25.72   & 23.78   &  7.52   \\
30   &  20.5  &  9.5  &  0 &  33.96   & 30.62   &  9.83   \\
50   &  22.5  &  9.5  &  0 &  35.34   & 31.54   & 10.74   \\
70   &  20.5  &  9.5  & 2.5&  23.10   & 20.47   & 11.6   \\
100  &  20.5  &  9.5  &  3 &  18.70   & 16.79   & 10.2    \\
200  &  20.5  &  9.5  &  6 &   8.58   &  8.07   &  5.9    \\
300  &  20.5  &  9.5  & 10 &   4.28   &  4.10   &  4.4    \\
400  &  22.5  &  9.5  & 12 &   2.84   &  2.74   &  3.34   \\
500  &  22.5  &  9.5  & 15 &   1.83   &  1.78   &  2.83   \\
1000 &  17.5  & 11.5  & 22 &   0.58   &  0.57   &  1.62   \\
\hline 
\multicolumn{7}{c}{$g = 100$} \\ \hline
10   &  19.5  &  9.5  &  0 &   60.94 &   44.24  &  27.40   \\
20   &  19.5  &  8.5  &  0 &  257.30 &  151.04  &  41.30   \\
30   &  19.5  &  8.5  &  0 &  335.08 &  183.08  &  45.18   \\
40   &  19.5  &  9.5  &  0 &  351.34 &  188.54  &  46.34   \\
50   &  19.5  &  8.5  &  0 &  344.52 &  186.06  &  45.99   \\
70   &  20.5  &  8.5  &  0 &  308.92 &  173.04  &  43.99   \\
100  &  20.5  &  8.5  &  0 &  255.74 &  152.88  &  40.22   \\
200  &  20.5  &  9.5  &  0 &  153.80 &  105.88  &  31.16   \\
500  &  20.5  &  9.5  &  5 &   43.06 &   33.51  &  22.20   \\
700  &  20.5  &  8.5  &  5 &   25.15 &   20.64  &  17.93   \\
1000 &  20.5  &  8.5  & 10 &   14.53 &   12.71  &  12.56   \\
2000 &  20.5  & 10.5  & 15 &    5.04 &    4.74  &   5.90   \\
3000 &  19.5  &  9.5  & 18 &    2.72 &    2.61  &   3.80   \\
\end{tabular}

\end{table}

\narrowtext

\noindent
the average velocity of the
WIMP's as a function of $\theta$ together with the average velocities
for the case of a transparent Earth.  Note that for all angles, the
average values of the distributions are lower than in the transparent
case meaning that, due to the interaction, the particles lose velocity
to some degree. In particular, those distributions when $\theta$
approaches 180$^{\rm o}$ would yield the largest effect.

Figures~\ref{fig3} and \ref{fig4} indicate that the passage of the
WIMP's through the Earth alters their velocity distributions. This
phenomenon and the net change in the number density of particles for
rear angles modify the differential rate with $\theta$ and are
therefore responsible for the effect of diurnal modulation.

Figure~\ref{fig5} illustrates the changes produced in the differential
energy spectra. In general, low masses are inefficient in depositing
energy in the detector because their momentum transfer to recoiling
nuclei is smaller, therefo-

\newpage

~
\vspace{15.82cm}

\noindent
re their signal decreases rapidly with
energy. Signals at high energies are only achievable by high-speed
WIMP's, but WIMPS with such velocities ($\sim 800$~km~s$^{-1}$) are
scarce. High masses, on the other hand, can deposit larger amounts of
energy. But, their number density decreases as the mass of the
candidates increases because of the constant halo density that galactic
dynamics requires.  This results in an overall signal decrease with
WIMP mass at high masses and a different slope on the graph.

The rather small variations in the velocity distributions at $\theta =
90^{\rm o}$ (day) and $\theta = 170^{\rm o}$ (night) are responsible
for the two sets of predicted energy spectra in Fig.~\ref{fig5} (solid
and dashed lines, respectively). Generally speaking, those WIMP's
scattering the Earth at rear angles have lower relative speeds and
therefore their signal decreases. However, this is only true for a
specfic range of masses. For masses 100~GeV and larger, the variations
in the velocity distributions change the slopes of the spectra
generating crossing points at definite energies. These crossings
energies, $E_{cross}$, occur at 1~keV for a mass of 100~GeV, 11~keV for
400~GeV, 20~keV for 1~TeV, and 22~keV for a 2~TeV WIMP mass.
Determining the minimum energy $E_{min} \, >\, E_{cross}$ and
integrating the energy spectra between $E_{min}$ and an upper limit
(say 50~keV) we {\em optimize} the predicted signal to obtain the
maximum variation in the detection rate above $E_{min}$.

These new calculations clearly show the energy $E_{cross}$ below which
scattering in the Earth increases the counting rate of WIMP's, and
above which the counting rate is decreased. This gives a new and more
sensitive technique to analyze experimental data for diurnal
modulations than just using the portion of the spectrum above
$E_{cross}$.

The variation of the differential rate with $\theta$ is all we need to
predict the WIMP's signal.  By combining this result with our knowledge
of $\theta$ as a function of time (Fig.~\ref{fig2}) one is able to look
for the characteristic signature in our detector due to WIMP's.

\section{Results and Conclusions}

The total expected number of counts per keV and kg~h in the range
$E_{min}$ to 50~keV is shown in Fig.~\ref{fig6}. In the graph, the
cross sections were enhanced a factor 100 with respect to $G_F^2$ in
order to make the diurnal modulation effect visible.  From the figure,
it can be clearly seen that the modulation effect is larger for masses
close to 50~GeV, whereas it is very small for masses greater than
500~GeV.  This is due to the asymmetry between the masses of the Earth
nuclei and the WIMP masses.  If the mass of the latter is many times
larger than that of the nuclei in the Earth, the WIMP continues its
trajectory almost unaffected by the interaction and no modulation is
expected.  For masses close to the average mass of nuclei in the Earth,
the WIMP may suffer a large deflection and the scattering angle can be
significantly different from zero.  In this case, we predict a clear
modulation effect.

To quantify the variations in Fig.~\ref{fig6}, we define the modulation
amplitude, $\delta$, as the ratio of the difference between the maximum
and minimum of the total predicted rates (between $E_{min}$  -- 50~keV)
with respect to the maximum rate,

\begin{equation}
   \delta \; = \; 100\, \frac{R_{max} - R_{min}}{R_{max}} \;.
\end{equation}

The behavior of the modulation amplitude, $\delta$, as a function of
the WIMP mass for different coupling constants, $g$, is shown in
Fig.~\ref{fig7} (the actual values are given in Table~\ref{t1}). In the
graph only data points for which modulation is clearly seen are shown.
Table~\ref{t1} also shows the LST at which the maxima and minima are
predicted.

\vspace{-0.7cm}

\begin{figure}[t]
\centering
\epsfxsize=9.0truecm
\epsfysize=6.8truecm
\epsffile{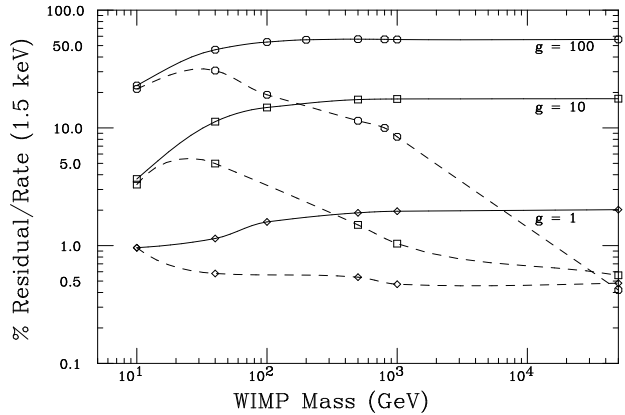}
\caption{Previous predictions of Ref.~\protect\cite{juanD} (Fig. 9)
in solid lines and new values obtained after the code was modified
(dashed lines).\label{fig8}}
\end{figure}

We notice that for the coupling constants considered here, the greater
the mass of the candidate, the smaller is the modulation expected. This
decrease of the modulation amplitude moves towards the lower masses as
we decrease the coupling constant.

The results of the present calculations are in disagreement with those
of Refs.~\cite{juanD,garciaD} where no degradation of the modulation
amplitude is predicted for increasing masses. A cross-check of the
calculations revealed that in all previous results a transformation of
the scattering angle from the center of mass to the laboratory frame
was missing.  When the correction is included the modulation amplitudes
decrease for masses larger than $\sim 50$ GeV bringing the results into
agreement.

Figure~\ref{fig8} summarizes the differences in predictions at 1.5~keV
before and after the correction.  This figure is analogous to Fig. 9
in Ref.~\cite{juanD}.  Notice that, since some of the new values of the
relative residual are negative,
 
\noindent 
Fig.~\ref{fig8} shows their {\em absolute values}.  At some point, the
spectrum of the high band is larger than the low band and this produces
a negative value of the residual.  Only three interaction strengths are
considered in the figure just to exhibit the modifications to the
original and to show the behavior of the relative residuals with the
mass of the WIMP's.  Now the new values show a behavior with WIMP's
mass similar to Fig.~\ref{fig7}.

In view of these new results, we conclude that a larger amount of data
than previously estimated is needed for the sensitivity of the diurnal
modulation method to exceed that of the conventional signal-to-noise.
The minimum modulation amplitude that can be detected or excluded
decreases, as is the case for any search for a periodic modulation, as
roughly the square root of the volume of data. For collected statistics
of $\sim$ 500~kg~day, and typical background levels of
2--3~counts/keV~kg~day~\cite{taup95} this translates in a sensitivity
of $\delta >$ 2\%, which would be produced by particles (see Fig.
\ref{fig7}) that are already excluded using the conventional
signal-to-noise method (for Dirac neutrinos, $g=1$, masses larger
than 26~GeV and up to 4.7~TeV have already been ruled out by
direct-detection experiments~\cite{H-M}). With these figures, at least
20~kg~yr of data are required for the diurnal modulation method to be
sensitive to $\delta \approx$ 0.5\% needed to extend the WIMP search
into the cosmological interesting region. It is worth noticing that
though the calculations were conducted assuming vector interactions
between the candidates and the Earth constituents our results also
apply to dark matter particles with scalar interactions.

The other alternative to identify a WIMP candidate using a
semiconductor detector is to look for {\em annual modulation} of the
signal.  Now the average velocity of the WIMP's impinging on the Earth
is what changes throughout one period (a year). Since this variation is
independent of the WIMP's mass, changes in the rate should be present
for any candidate.  At the present time, an exhaustive data analysis of
the DEMOS collaboration is being carried looking for annual modulations
of the signal in different parts of the spectrum.

\acknowledgements{The authors would like to express their appreciation
to Professor F.~T.~Avignone for his help and guidance throughout the
doctoral research of F.H. from which this work resulted.}

\end{document}